\documentstyle[twocolumn,aps]{revtex}
\begin{document}

\newcommand{\beq}{\begin{equation}}
\newcommand{\eeq}{\end{equation}}
\newcommand{\bea}{\begin{eqnarray}}
\newcommand{\eea}{\end{eqnarray}}
\newcommand{\ba}{\begin{array}}
\newcommand{\ea}{\end{array}}
\newcommand{\om}{(\omega )}
\newcommand{\bef}{\begin{figure}}
\newcommand{\eef}{\end{figure}}
\newcommand{\leg}[1]{\caption{\protect\rm{\protect\footnotesize{#1}}}}

\newcommand{\ew}[1]{\langle{#1}\rangle}
\newcommand{\be}[1]{\mid\!{#1}\!\mid}
\newcommand{\no}{\nonumber}
\newcommand{\etal}{{\em et~al }}
\newcommand{\geff}{g_{\mbox{\it{\scriptsize{eff}}}}}
\newcommand{\da}[1]{{#1}^\dagger}
\newcommand{\cf}{{\it cf.\/}\ }
\newcommand{\ie}{{\it i.e.\/}\ }
\newcommand{\eg}{{\it e.g.\/}\ }

\title{Contextual objectivity : 
a realistic interpretation of quantum mechanics.}
\author{Philippe Grangier}
\address{Laboratoire Charles Fabry, Institut d'Optique Th\'eorique et
Appliqu\'ee, 
F-91403 Orsay, France}

\maketitle

\begin{abstract}

An attempt is made to formulate quantum mechanics (QM) in physical rather than
in mathematical terms. It is argued that the appropriate conceptual framework
for QM is ``contextual objectivity", which includes an objective definition of
the quantum state. This point of view sheds  new light on
topics such as the reduction postulate and the quantum measurement process.


\end{abstract}

\section{Introduction}

We attempt here to define a physical conceptual framework
adapted to the interpretation of the usual formalism of quantum mechanics. We do not
make any changes to the mathematics of this formalism, neither do we
introduce new equations. Our aim is rather to present a physical interpretation
of the formalism in
such a way that the so-called ``difficulties" of QM, such as the reduction
postulate, or the measurement process, can be seen from another side.
Our approach is different from the one using ``consistent (or decoherent) histories";
it carefully avoids ``multiple worlds", and it has nothing to do with any
revival of ``hidden variable theories". It should be better to consider it as a
``neo-copenhagian" point of view, in the sense that 
it formalizes a distinction between the classical and quantum worlds. 
However, contrary to the copenhagian dogma,
a central point in our approach will be to give
an ``objective reality" to the quantum state of a physical system, 
in a sense which is developed below.

\section{A physical definition of the quantum state}
\label{def}

In their celebrated paper written in 1935, Einstein, Podolsky, and Rosen
made the
following statement \cite{epr}:
``If without in any way disturbing a system we can predict with certainty
(i.e.,
with a probability equal to unity) the values of a set of physical
quantities, then
objective reality can be attached to this set of physical quantities".
It is known that this rather ``metaphysical" definition resulted in a lot of
questioning and arguing. However, this will be our starting point. We will
slightly modify
this statement in order to give our definition of the quantum 
state\footnote{Throughout this paper ``state" means ``pure state". 
Mixed states, when needed, will be called ``statistical mixtures" \cite{cct}.} as follows.

{\bf The quantum state of a physical system is defined by the values of a
set of
physical quantities, which can be predicted with certainty and measured
repeatedly without perturbing in any way the system. This set of quantities must be
complete in the sense that the value of any other quantity which satisfies
the same criteria is a function of these values.}

We will first make some comments about this definition, then draw some
consequences from it

Comment 1 : The definition is clearly in agreement with the usual formalism of
QM, as it can be seen when using the notion of ``complete set of
commuting observables" (CSCO) \cite{cct}. A quantum state is specified by the
ensemble of eigenvalues corresponding to a CSCO, which can obviously be
measured repeatedly without perturbing in any way the system.
Actually, any physical quantity that satisfies the definition can be
expressed as a
function of the CSCO observables. This ``completeness" condition is essential
to warrant a unique correspondence between the state and the set \cite{jlb}.

Comment 2 : Since we consider the usual QM formalism as said above, the
quantum state is mathematically described by a vector in an Hilbert space, and
it evolves in time according to Schroedinger's equation. It it clear that unitary
evolution from Schroedinger's equation transforms a state which satisfies our
definition into a similar state, associated with a different set of physical
quantities, corresponding to new well-defined measurements 
(that may however be not easy to perform).

Comment 3 : As an example,
a coherent state $| \alpha \rangle$ of the harmonic oscillator can be in a
deterministic way displaced by  $(- \alpha)$ \cite{cct}. After that operation,
a measurement of the photon number $\hat n$ gives $n=0$ with
probability one. A similar reasoning applies to an evolving gaussian wave
packet: if we catch it in an appropriate harmonic potential, and we measure the
energy, we find  the ground state energy with probability one. Generally speaking, our
definition simply uses the fact that a pure state is always a joint
eigenstate of a set of commuting operators.

Comment 4 : An obvious but fundamental point is that given a quantum
state, not all possible physical quantities can be predicted with certainty, but
only a subset of them. That the subset is the largest 
possible set of independant quantities is just the
definition of a CSCO. In the definition, this appears in the statement ``measured
repeatedly without perturbing in any way the system" : if the subset is not a
CSCO, a measurement will generally change the state. How to deal with
physical observables which do not commute with those of the CSCO is a
crucial point, which will be discussed in section \ref{meas}.

Comment 5 : In order to appreciate the difference between our definition and
the EPR reasoning, one should consider what ``can be predicted with certainty"
in a two-particle EPR state, such as a singlet state. In \cite{epr}
EPR were considering measurements on separated
subsystems, where the {\it conditional} probability of
a measurement result on one particle,
given a measurement done on the other particle, may take the value one.
However, the {\it overall} result of both measurements remains inherently uncertain,
simply because measurements on separated subsystems are not part of the
CSCO defining the EPR state (see also section \ref{meas}). What actually gives a
certain and reproducible result is a joint ``Bell measurement" on the pair of
particles \cite{special}. An EPR state thus perfectly fits within our definition.
Moreover, we may define quantum non-separability as the impossibility
to define the quantum state (according to the definition given above) for a sub-part
of a system which is globally in a spatially extended entangled state.

Consequence 1 : Though this may sound metaphysical, our definition
implies that some ``objectivity" can be attached to the quantum state, in a
sense
which is remarkably close to the one used by Einstein, Podolsky and
Rosen. In particular, a quantum state as defined above is clearly
independent of
the observer. Also, it is well known from statistical physics or quantum
information theory that a pure state has zero entropy. This clearly fits
with a set
of perfectly predictable and observer-independent properties.

Consequence 2 : Our definition can be given a meaning only by assuming that
the sentence ``predicted with certainty and measured repeatedly" itself has a
meaning, i.e., that we are in a context where this predicate corresponds to a
well defined action (i.e. a possible experiment). This is why we call our
point of view ``contextual objectivity": the quantum state does have an
objective
existence, but its definition is {\bf inferred} from observations which are
made
{\bf at the macroscopic level}. We point out that there is no need to refer to
``observer's consciousness" or anything like that: all what we need is
simply the
the usual classical world, as the place where the measurements are made, and
where their results can be recorded by any (conscious or unconscious) observer.

Consequence 3 : According to the above observations, the state of the
system is
conditioned by the external classical world, which, from now on, we will
call the
``environment". Therefore, we can use an inference principle: from
an appropriate observation in the environment, one can infer the quantum state
of a system. This point is crucial in what follows, because it is for
instance in contradistinction with a ``multiple world" point of view: in
our approach,
there is only one environment, and the ``reality" does not only develop
upwards
by ``ramification" of the quantum state, but also downwards by inferring
microscopic states from macroscopic observations.

Consequence 4 : A fundamental consequence of all the above is that what we
call the quantum state {\bf cannot} involve different ``statuses" of the
environment.
Different statuses of the environment can occur only if they are associated
with different quantum states. This may include the case where the status of the
environment is not known, and is described by using classical statistics.
This corresponds to the usual ``statistical mixtures" in the density matrix
formalism. It should be pointed out that a statistical
mixture, {\it contrary to a pure state}, is ``contingent" (\ie observer-dependant). 
In other terms, a statistical mixture is always associated to a (classical) missing
information, which may be known by somebody else. On the other hand, nobody can
know more about a pure state, than what is given in its definition \cite{dm}. 

We will argue below that within the framework defined above, there is no need
to add a ``measurement postulate": this postulate is already implied by the
very
definition of the quantum state.

\section{Measurements on a quantum system}
\label{meas}

We now come to the measurement of a physical quantity which does not pertain
to the set
allowed by the definition of the state (\ie the CSCO), and
therefore which cannot be predicted with certainty. In a first approach, we
will
adopt the usual ``decoherence" point of view \cite{special}, which is
basically an attempt to
calculate what is going on during the measurement, using the initial quantum
state and Schroedinger's equation. Then we will criticize this approach and
introduce our point of view.

In the decoherence approach, the initial steps of the measurement of a
physical quantity for a system S1 can be described by using
Schroedinger's equation, as the interaction of two (or several) systems S1,
S2, ... .
As a result of this interaction, S1 is generally no longer in a well
defined state,
i.e. no physical quantity relevant to S1 only is predictable with
certainty. The
set of fully predictable physical quantities, which still exists by
definition of the
state, will rather correspond to joint measurements of S1 and all the other
systems it has interacted with.

Remaining within the decoherence approach, the evolution of the ``growing"
system will
at some point involve different
possible values of environment variables, such as the position of a needle.
Then several things happens. First, the
"fully predictable quantities" which are still attached to the global state
become
useless in practice, because the corresponding joint measurements become
unpracticable. Second, the calculation can be pushed further by
concentrating on
the system itself. This leads eventually to different possible quantum
states for
the system, each of which is associated with a macroscopically different
environment, and occurs with a probability which is given by the calculation.
There are several ways to look at these alternative possibilities:

(A) one may postulate that after the measurement the system is ``projected"
onto one
possibility only, corresponding to the observed environment (usual ``reduction
of the wave packet" postulate)

(B) one may assume that all possibilities keep on going, but that there are
continuous
``branching" of universes, so that we can see one branch only, i.e. the one
where we
live (``many-worlds" interpretation)

(C) one may assume that the alternative possibilities describe a
statistical ensemble of
identically prepared systems. In accordance with the usual statistical
approach,
there is some ``missing information" about the state of the system,
corresponding to the
fact that the density matrix of the system alone is a statistical mixture.
This
information is supposed to be written in the environment.

We list below some difficulties associated with these different points of
views,
then we draw a conclusion.

(i) The point of view (A) has the problem of ``terminating" the unitary
evolution
of the state (according to Schroedinger's equation) by an ``extra" non-unitary
process (a projection). The question is then how to relate these two radically
different types of evolution.

(ii) The point of view (B) does not explain why only one result is observed in
any given experiment: in (A), the uniqueness of the result is postulated,
in (B) all
possible results are still there, and in (C) this question is ignored and
considered
to be irrelevant.
The usual answer in (B), which says that only the branch ``where I am"
actually
matters, is very puzzling, because the existence of other branches ``where
I am
not" looks more as an artifact than a description of the physical world.

(iii) In the point of view (C), that we believe to be the most widely
accepted at present,
the existence of a missing information written in the environment is
postulated
rather than demonstrated. Actually, it can be argued that the choice of one
(and
only one) measurement result ``must happen" at some point; then (C) seems to
require an implicit use of either point of view (A) or (B).

More generally, all that can be done by the decoherence point of view is to
create a ``growing quantum state" as more and more sub-systems are coupled to
the initial system S1. As far as we can see, this cannot warrant that a
classical-
looking environment will be recovered at the end. On the other hand, one must
remember our initial argument that the environment was actually a
 pre-requisite for the definition of the quantum state.

We are thus lead to our main conclusion, which is simply that {\bf the
uniqueness of
the measurement's result is guaranteed by the uniqueness of the environment,
which is actually, as it was said from the beginning, a necessary condition
for
the very definition of the quantum state}.

Comment 1 : This may hurt the intuitive feeling that the measurement should
either ``reveal" a pre-existing state of S1 (naive ``classical" view), or
``drive" S1
from an initial to a final state, including the environment (naive ``quantum"
view). Actually, for quantities which are not in the initial CSCO, there is in
general {\bf no} pre-existing quantum state (i.e. well-defined properties)
of S1, to be
revealed at the end of the measurement. On the other hand, S1 is not ``driven"
somewhere: according to the QM formalism, S1 has become intricated with the
entire environment, and it is only as a consequence of the procedure
which allows one to define a quantum state, that such a state is eventually
observed.
The ``contextual" essence of QM lies in the compatibility
between the ``downstream" definition of a quantum
state, and the ``upstream" evolution given by the decoherence calculation.

Comment 2 : In our approach,
{\bf the boundary between quantum and classical mechanics results from the
fact
that the quantum state is defined in terms of the observed environment, and
thus
cannot include this environment}. If the evolution of the system leads to an
``entanglement with the environment", then the global state becomes irrelevant
(because the corresponding fully predictable observables are out of reach),
and the
system's state is no longer defined. It appears then that the system is in a
statistical mixture of states, where the ``missing information" can be read
out in
the environment. The decoherence approach allows one to calculate the
corresponding diagonal density matrix, but it is unable to ``terminate" the
calculation, by extracting only one result. The ``interpretative" step that
is required
at this point is more than a matter of convenience related to the fact
that the global state is too complicated to be followed.
Our claim is that {\it this interpretative step is required from the beginning}, because
it is precisely the one which is required in order to define
the quantum state (see section \ref{def}). In
practice, our conclusion agrees quite well with the point of view (C)
introduced
above, which is now built inside our very definition of the quantum state, and
does not require an implicit use of either (A) or (B).

Comment 3 : A well known question is then: how to separate the ``system"
from the ``environment" ? Actually, defining the environment is not ambiguous,
since it is the point from which the quantum state is defined (i.e. the
classical
world where we are). The definition of the system is more
subtle, and it depends on the precise physical system which is studied. The
existence of this (very flexible) boundary, far from being a weakness of
QM, is
actually a strength. In particular, this opens the way to ``mesoscopic quantum
states", which are one of the main challenges of today's quantum mechanics.

Comment 4 : Our view is consistent with the idea that QM is a non-
deterministic theory. Though the evolution of the quantum state is
deterministic, non-determinism is due
to the fact that in general (\ie each time a non-CSCO observable is
measured) the measurement redefines the state,
taking into account the contextual nature of the theory.
Similarly, our approach does not change anything to quantum non-separability:
a quantum state is generally non-separable, and may violate Bell's
inequalities.
Despite our initial use of the EPR wording, our point of view thus
differs from EPR on two central issues:

(i) not all physical quantities are fully predictable, only a subset of them
(Heisenberg's inequalities)

(ii) the quantum state has an objective reality, but it is non-separable
(Bell's inequalities).

Non-separability is probably the most original feature of quantum mechanics,
and directly contradicts ``local realism", which is the view expressed
mathematically in the hypothesis leading to Bell's inequalities.
On the other hand, there is no contradiction with relativistic causality,
and not even with ``naive" realism, according to our definition of the quantum state. 

Comment 5 : Do we add anything new with respect to the ``reduction of the
wave packet postulate" ? Again, the difference is that this postulate
is no longer necessary, but is built inside our definition of the quantum
state:
from the beginning, a quantum state is (objectively!) defined with respect to
observations which are made within the environment. It is just a
consequence of this
definition that new observations in a changing environment lead to a new - but
again uniquely defined -  quantum state.

Comment 6 : Clearly our very definition of a quantum state relies on something
external to it, as was often advocated by Bohr. However, we do not accept
to infer that a quantum state is merely ``subjective knowledge". We
consider that our definition of the quantum state is objective, in a sense
that has
been discussed above. In our view, {\it a quantum state gives the best
possible
objective description of a suitably isolated subsystem of the physical world}
(what is a suitable isolation is defined by QM as being
``decoherence-free"). On
the other hand, we note that our point of view makes it very difficult to
speak about
anything like a ``quantum state of the universe".

Comment 7 : Most interpretations of QM, including the present one,
agree on the fact that the quantum structure of the microscopic world explains
the features of the classical structure of the macroscopic world. But another
question is  in which sense it is possible to ``construct" or ``deduce"
classical
physics from quantum physics. There are at least two answers to this question.
The first one is simply to say that
the purpose of QM is to calculate correlation functions between successive
measurement events. This is a very ``minimalistic" approach from a
conceptual point of view,
but it is extremely useful both for teaching and for using QM. Our approach
is fully compatible
with this ``correlation function" approach: in some sense, we are only
trying to put words
around it, in a way as consistent as possible. Another answer is to say that
the classical world can be explicitly deduced from QM by using \eg coarse
graining and
decoherent histories. However, this point of view leads also to the idea of
the ``wave
function of the universe", including an infinite infinity of branches
corresponding to all possible results of all possible ``measurements". We are
clearly very doubtful about these approaches, precisely because the
formulation
of QM is contextual, that is, depends at its start from what it should
``explain" at
its end, i.e., the observed environment. As said above, our ``contextual
objectivity" point of view is that QM is the best possible objective
description of
a ``suitably isolated" (or ``decoherence-free") {\bf subsystem}. This is
already a lot
in terms of predictive power, but if carried out of this realm, QM does
turn into
a ``smoky dragon", which tries in vain to capture its origin.

\section{Conclusion}

With the general goal to provide a ``status report" rather than any final
argument, we have presented an attempt to reconcile the still prevailing
Copenhagian point of view about QM, with the more modern ``decoherence"
points of view. A brief summary of the main our main statements is presented here :

(1) Measurements allow one to define the quantum state of a physical system
in an objective and observer-independant way
(the new point here is that we apply this ``classical" statement to QM as well).

(2) Given a quantum state, only a {\it subset} of all physical quantities is fully predictable
(on the other hand, given a classical state, {\it all} physical quantities can
be known; denying this possibility {\it does not} impede our objective
definition of the quantum state).

(3) A quantum state may be non-separable : the physical properties of a sub-system are then
among the non-predictable ones, even though the state of the whole system is perfecty defined
(on the other hand, the classical state of a compound system can always be defined by
cutting it in parts, and defining the state of each part; again, denying 
this possibility {\it does not} impede our objective definition of the quantum state).

(4) If one measures a physical property which is not among the predictable ones,
the initial system becomes a (non predictable) sub-system in the measurement apparatus,
and the state is reset by looking at the measurement result
(this brings back to step one).

Our approach differs from the orthodox Copenhagian approach, by insisting
upon the objective character of the quantum state. It also differs from the
standard decoherence approaches, by insisting upon the fact that the boundary
between classical and quantum mechanics cannot be ``erased", since a
quantum state ``involving the environment" cannot be consistently defined.

Both aspects are contained in the definition of a quantum state from the values
of a set of physical quantities which can be predicted with certainty and
measured repeatedly without perturbing in any way a system.

\section*{Acknowledgments}

Though I am sole responsible for all the wrong statements that may appear in this paper,
its content owes a lot to many discussions with Franck Laloe, who consistently
pointed out to me weaknesses in the standard decoherence approach.
Many thanks are due to Jean-Louis Basdevant for judicious comments and corrections.


\begin{references}
\bibitem{dm}
In the line of the present approach, the most natural way 
for specifying a statistical mixture is by using the eigenvalues and eigenvectors
of the density matrix. The eigenvalues clearly appear
as the probabilities of different (orthogonal) states represented by the eigenvectors, 
as it would be the case classically. 
When the density matrix has degenerate eigenvalues, the eigenvectors basis is not unique.
In our approach, this simply means that not enough (classical) information is available
to specify the system's state.
\end{references}
\end{document}